\documentclass{PoS}

\usepackage{epsf}
\usepackage{psfig}
\usepackage{epsfig}
\usepackage{lscape}

\title{3C84, BL Lac.
Earth based VLBI test for the RADIOASTRON project 
}

\ShortTitle{Software Correlator}

\author{A. Chuprikov\\
        Astro Space Center of P.\ N.\ Lebedev Physical Institute of
Russian Academy of Sciences,Profsoyuznaya 84 / 32, 117997,
Moscow, Russia\\
        E-mail: \email{achupr@asc.rssi.ru}}

\author{I. Guirin, A. Chibisov, V. Kostenko, Y. Kovalev\\
        Astro Space Center of P.\ N.\ Lebedev Physical Institute of
Russian Academy of Sciences,Profsoyuznaya 84 / 32, 117997,
Moscow, Russia\\
        E-mail: \email{igirin@asc.rssi.ru}}

\author{D. Graham, A. Lobanov\\
        Max Plank Institute for Radio Astronomy, Bonn, Germany\\
        E-mail: \email{p062gra@mpifr-bonn.mpg.de}}

\author{G. Giovannini\\
        Istituto di Radioastronomia di Bologna, Bologna, Italy\\
        E-mail: \email{ggiovann@ira.inaf.it}}

\abstract{Results of processing of data of a VLBI experiment titled RAPL01 are presented. 
These VLBI observations were made on 4th February, 2010 at  6.28 cm between the 100-m antenna of the 
Max Planck Institute (Effelsberg, Germany), Puschino 22-m antenna (Astro Space Center (ASC), Russia), 
and two 32-m antennas of the Istituto di Radioastronomia di Bologna (Bologna, Italy) in Noto and Medicina. 
2 well-known sources, 3C84 (0316+413), and BL Lac (2200+420) were included in the schedule of observations. 
Each of them was observed during 1 hour at all the stations. The Mark-5A registration system was used at 
3 European antennae. The alternative registration system known as RDR (RADIOASTRON Data Recorder) was 
used in Puschino. The Puschino data were recorded in format RDF (RADIOASTRON Data Format). 
Two standard recording modes designed as 128-4-1 (one bit), and 256-4-2 (two bit) were used in the experiment. 
All the Mark-5A data from European antennae were successfully converted into the RDF format. 
Then, the correlation function was  estimated at the ASC software correlator. A similar correlation function 
also was estimated at the Bonn correlator. The Bonn correlator reads Mark5A data, the RDF format was converted 
into Mark5B format before correlation. The goal of the experiment was to check the functioning and 
data analysis of the ground based radio telescopes for the RADIOASTRON SVLBI mission}

\FullConference{GAMOW-2010\\
		 August 23 - 27 2010\\
		 Odessa, Ukraine}

\begin{document}

\section{Results of data processing} 

The correlator estimates the visibility as a function of time and frequency. A phase of this complex function 
is shown in Figures 1, and 2.

\begin{figure}[h1]
\vskip2mm
\centerline{\psfig{figure=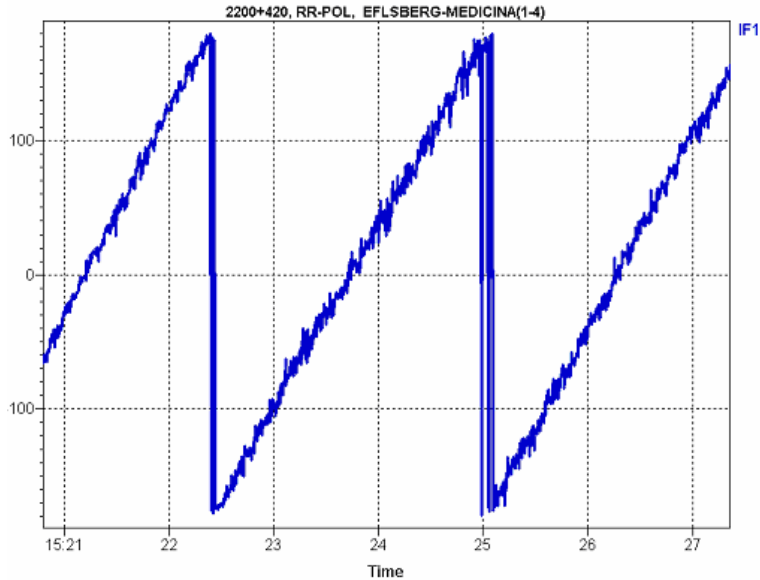,width=100truemm,angle=0,clip=}}

\caption{Visibility phase vs time for the Effelsberg-Medicina baseline. Scan no. 3 (2200+420). 
The maximum integration time is 2 minutes}
\label{fig:phase1}
\vskip2mm
\end{figure}

If a fringe rate (or fringe frequency) is still not compensated, there is a 
considerable phase slope. This may be due to several effects. For instance, it could be considerable if there 
is some instability of heterodyne at any antenna of a baseline. A slope restricts the coherence time which is 
equal to time interval corresponding to the phase drift of 180 degrees. 

\begin{figure}[h2]
\vskip2mm
\centerline{\psfig{figure=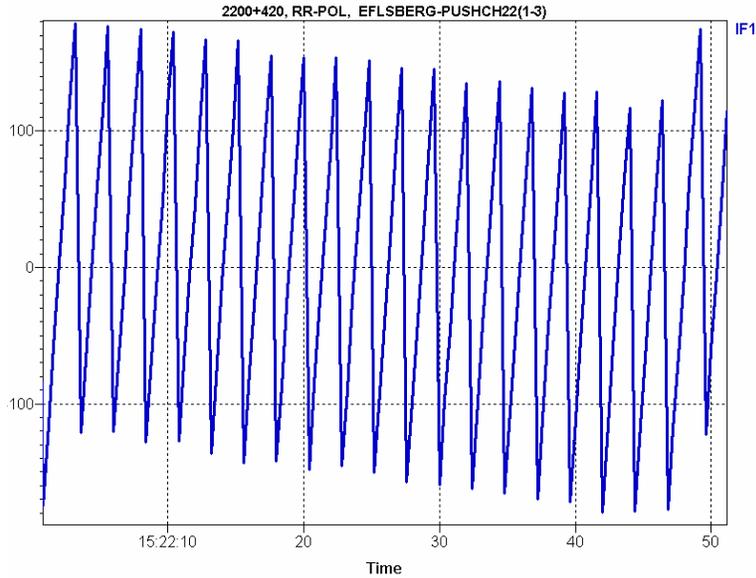,width=100truemm,angle=0,clip=}}

\caption{Visibility phase vs time for the Effelsberg-Puschino baseline. Scan no. 3 (2200+420). 
The maximum integration time is 0.3 seconds}
\label{fig:phase2}
\vskip2mm
\end{figure}

The correlator of ASC uses the special 
software module titled ARIADNA for estimation of delay for any interferometer baseline and any interval of 
observational time. Then, there is a possibility to compensate the residual delay as well as the residual 
fringe rate at the input of the correlator.

Values of  residuals with respect to the reference antenna 
in Effelsberg were the following : 
\begin{enumerate}
\item Residual delay values are :\\
$\bullet$ 21.10 +/- 0.05 mcs for the Puschino antenna\\ 
$\bullet$ -42.810 +/- 0.007 mcs for the Medicina antenna\\
$\bullet$ 12.520 +/- 0.005 mcs for the Noto antenna\\

\item Residual rate values are :\\
$\bullet$ 0.412 +/- 0.0005 Hz for the Puschino antenna\\
$\bullet$ 0.007 +/- 0.0006 Hz for the Medicina antenna\\
$\bullet$ -0.002 +/- 0.0005 Hz for the Noto antenna\\
\end{enumerate}

Values of residual delay and fringe rate were almost constant in time for each antenna during the observations. 
Thus, they could be compensated very easily inside the correlator. For example, the dependence of visibility 
phase on time for the second scan (source is 2200+420, lower frequency band, RR-polarization) is shown in the 
Figure 3.

\begin{figure}[h3]
\vskip2mm
\centerline{\psfig{figure=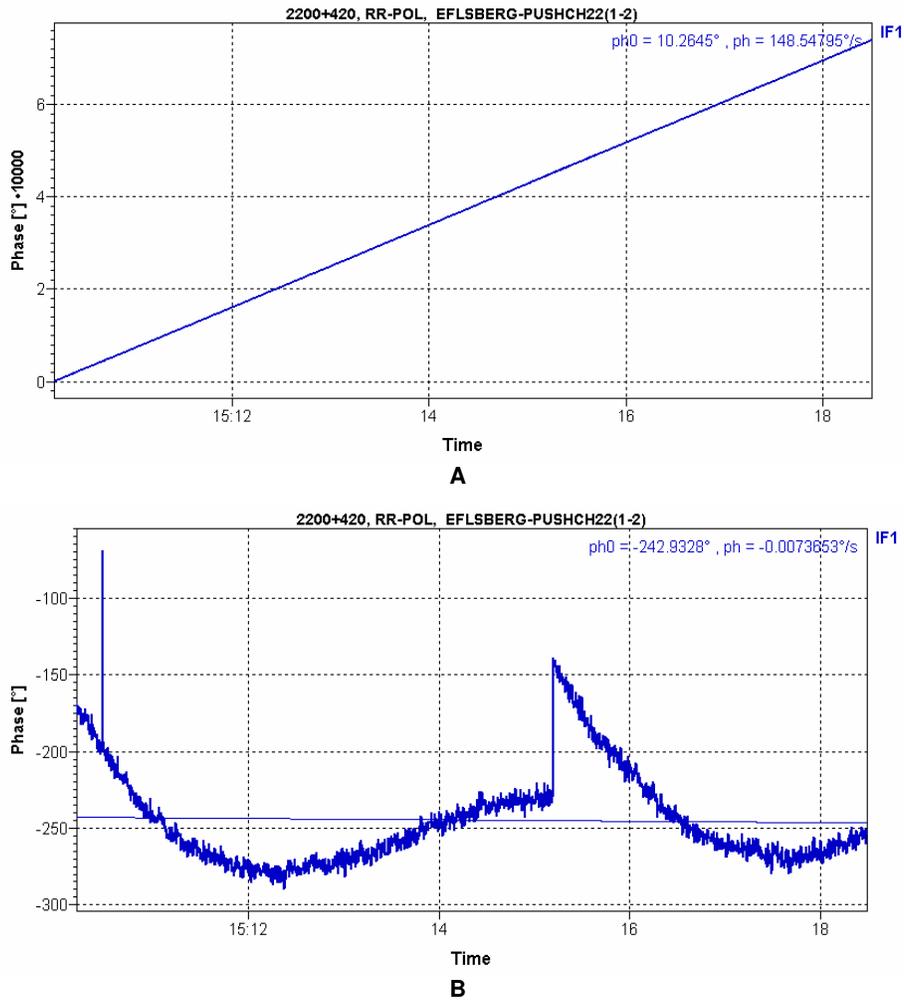,width=120truemm,angle=0,clip=}}

\caption{Visibility phase vs time for the Effelsberg-Puschino baseline BEFORE (A) and AFTER (B) 
the residual fringe rate compensation}
\label{fig:phase3}
\vskip2mm
\end{figure}

It is clear from Figure 3 that visibility phase values are between -280 and -140 degrees after the 
compensation. Such the residual phase oscillations are reliable and can be corrected during the secondary data 
processing. The value of residual fringe rate can be easily estimated from the phase slope value 
(148.54795 degrees per second). Then, the residual fringe rate value is equal to 0.4137. This is close to the 
value mentioned above (0.412).

The Signal-to-Noise Ratio (SNR) was measured for each baseline. The dependence of the Signal to RMS 
multiplied by 6 value on integration time for the data of scan no. 8 from the Effelsberg-Noto baseline and 
RR polarization is shown in Figure 4. 

\begin{figure}[h4]
\vskip2mm
\centerline{\psfig{figure=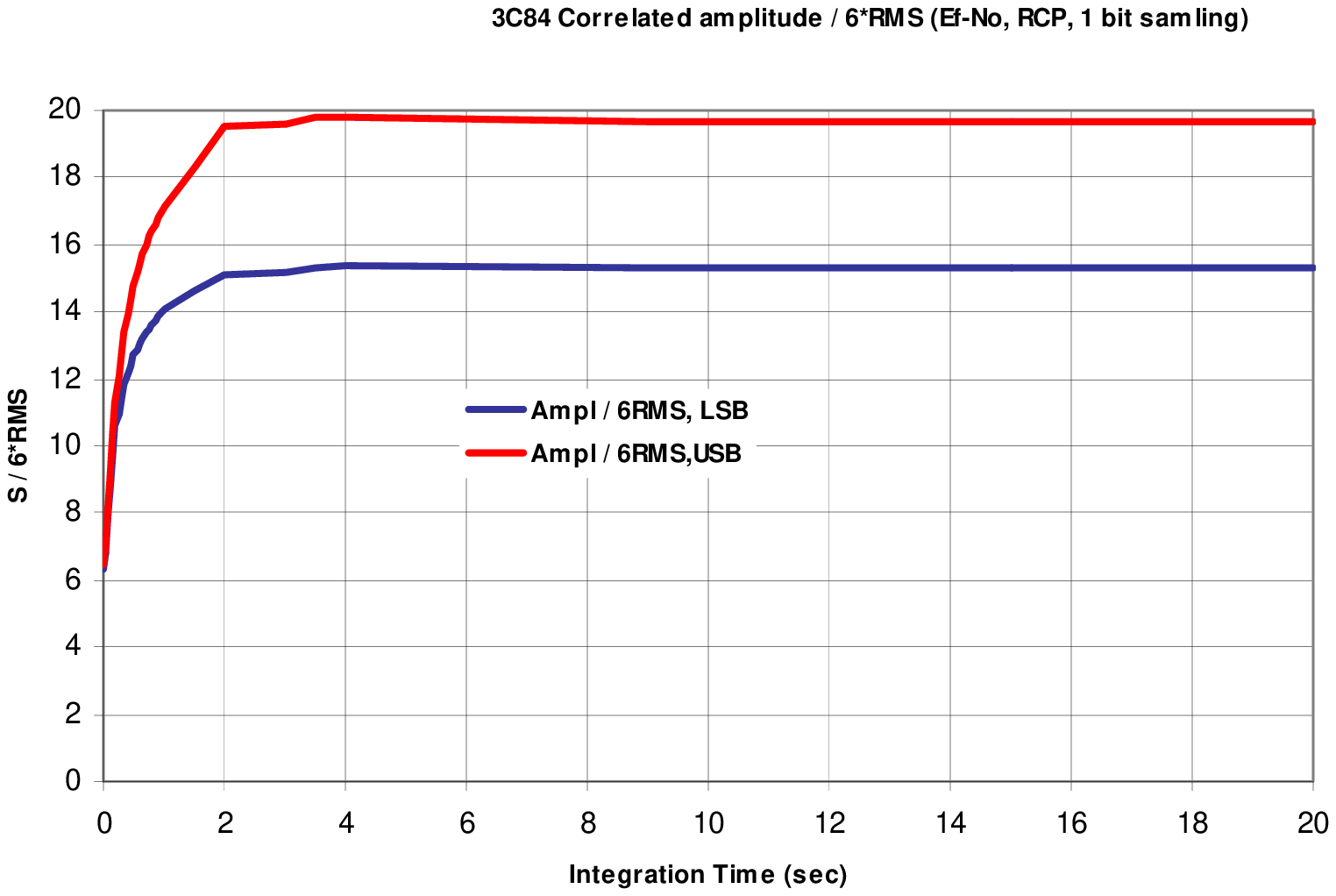,width=120truemm,angle=0,clip=}}

\caption{Signal-to-Noise ratio (SNR) vs the integration time}
\label{fig:snr4}
\vskip2mm
\end{figure}

It is clear from this figure, that the saturation of SNR takes place 
when the integration time is more than 2 seconds. Thus, there is the serious restriction for this parameter 
value even for European baselines, and its optimum value is about 0.3 second as mentioned above. 
There were observational data for the source titled 0316+413 (3C84) for all 4 antennae. A (u, v)-plane 
coverage (Figure 5) does not allow to reconstruct the source map. 

\begin{figure}[h5]
\vskip2mm
\centerline{\psfig{figure=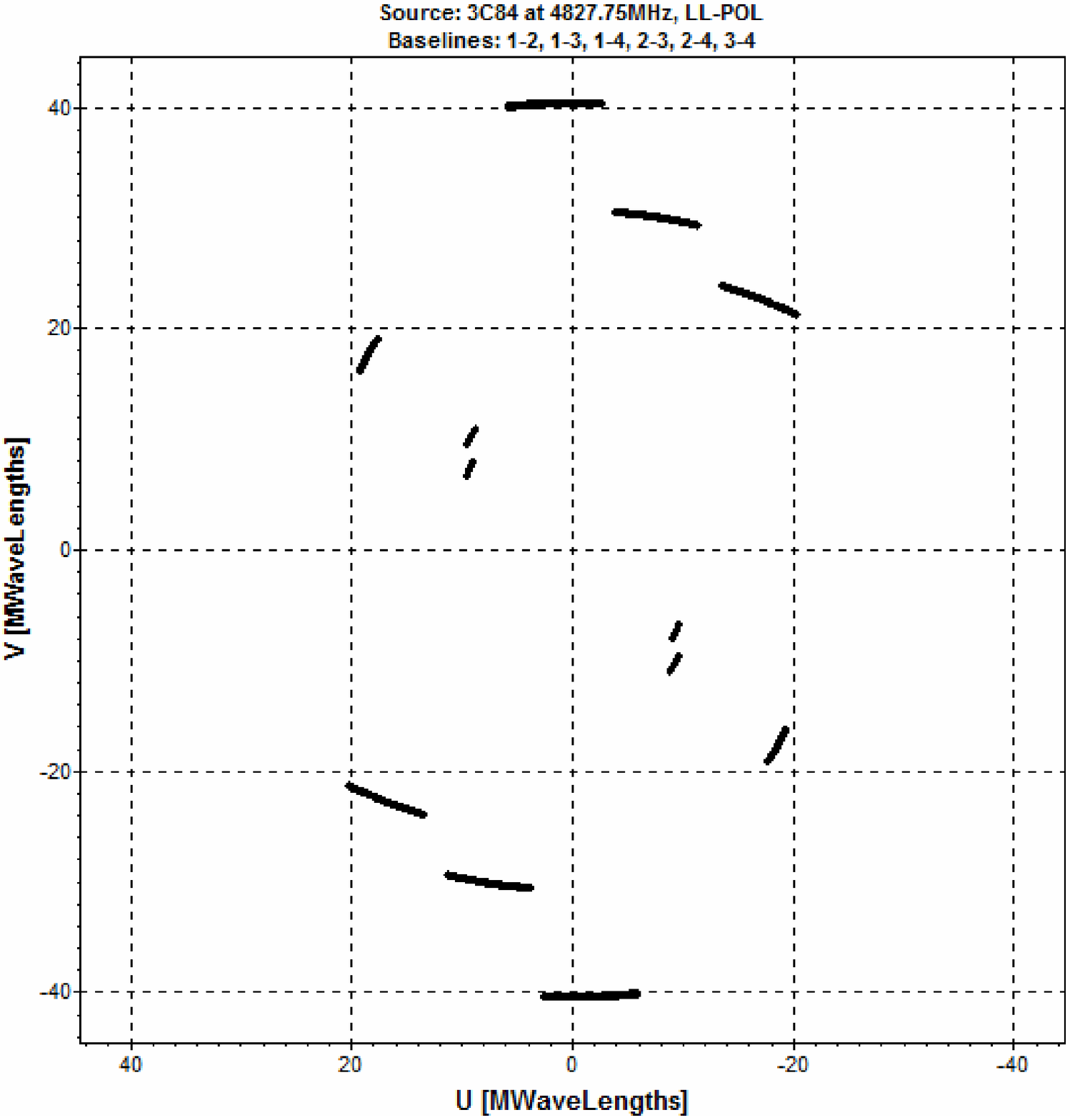,width=80truemm,angle=0,clip=}}

\caption{(u, v)-plane coverage for 3C84}
\label{fig:uv5}
\vskip2mm
\end{figure}

Estimation of the visible angular sizes of 
3C84  gives the value of about 11 mas (Figure 6). This is in consistent with the 3C84 properties available in 
literature (Asada, 2000, 2006).

\begin{figure}[h6]
\vskip2mm
\centerline{\psfig{figure=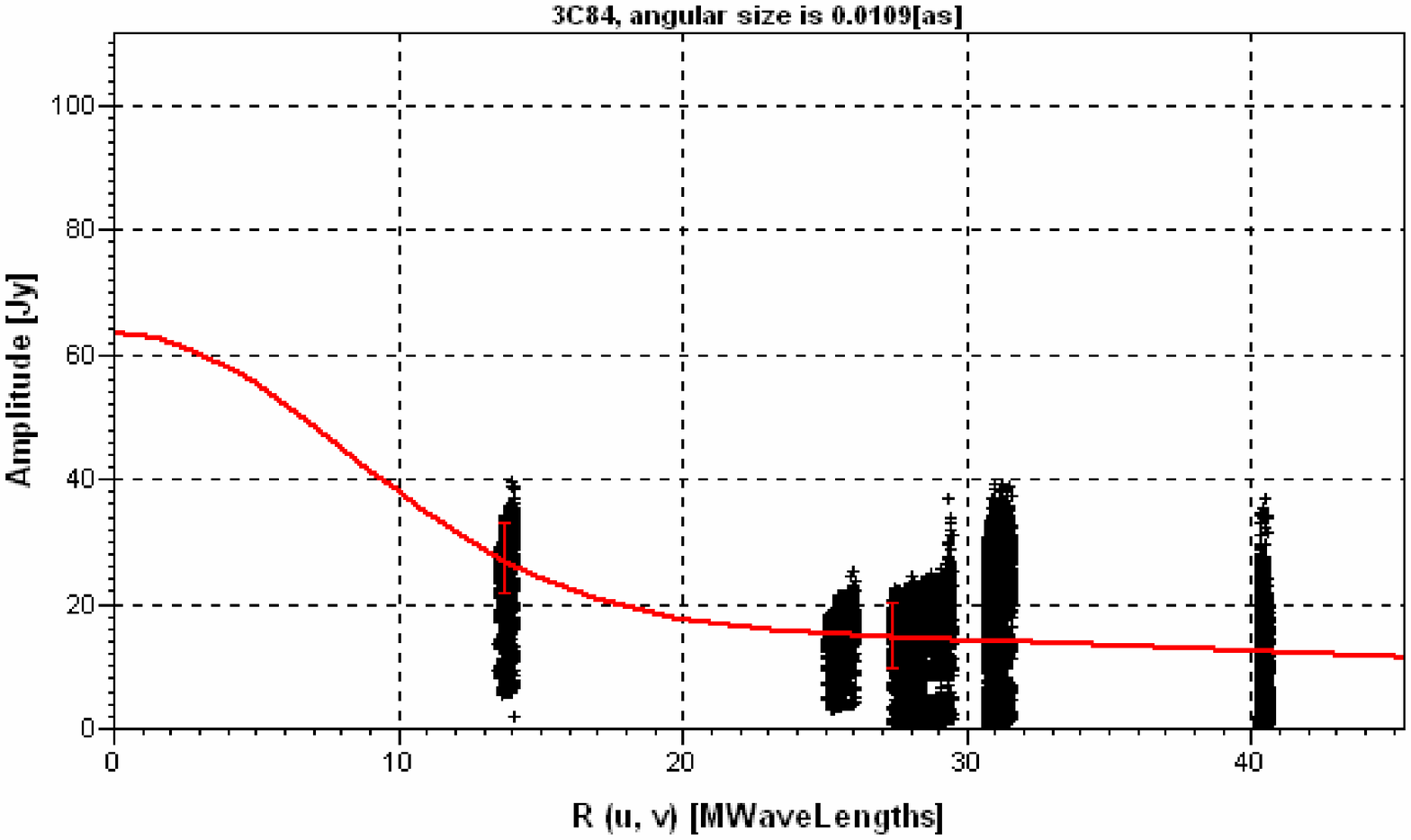,width=120truemm,angle=0,clip=}}

\caption{Visibility amplitude versus (u, v)-radius for 3C84}
\label{fig:avsr6}
\vskip2mm
\end{figure}

\section{Conclusions}

Thus, we could make the following conclusions :
\begin{enumerate}
\item The experiment RAPL01 demonstrates the possibility to convert the Mark-5A data into  RDF data. 
      Antennae with different registration systems could be successfully used for the RADIOASTRON mission\\

\item The integration time value is restricted by 1 second due to a high rate offset at Puschino antenna. 
      The successful estimation of the correlation function demonstrates the possibilities of the ASC software 
      correlator to compensate correctly the abnormally high values of residual delays and fringe rates\\

\item The data at the end of the ASC software correlator are relevant for the secondary processing\\

\item The calibration procedures of the software known as Astro Space Locator allow  
      reconstructing the visibility function\\

\item The (u, v)-plane coverage for the 3C84 is not sufficient to perform the source imaging. The value of 
      estimated source angular size is 11 mas. This value is consisted with the 3C84 properties available in 
      literature\\

\end{enumerate}

All the results presented in this paper are preliminary.\\
Procedures and techologies used during the VLBA data processing also could be very useful for 
processing of data of future Space VLBI mission titled RADIOASTRON.

\end{document}